\documentclass[twocolumn]{autart}    

\usepackage{graphicx}          
\usepackage{cite}                              

\usepackage{amsmath,amssymb}
\usepackage{graphicx}          
\usepackage[dvips]{epsfig}
\usepackage{bm}    
\usepackage{amssymb}
\usepackage{amsmath}

\usepackage{color}
\usepackage[T1,safe]{tipa}
\usepackage{esint}
\usepackage{bbm}
\usepackage{amsfonts}
\usepackage{mathrsfs}
\usepackage{cite}

\begin{document}

\begin{frontmatter}
\title{
Force Tracking in Cavity Optomechanics  with \\ a Two-Level Quantum System by Kalman Filtering
\thanksref{thk}}

\thanks[thk]{This work was supported by the National Natural Science Foundation of China under Grant 61873317, Grant 61828303, and Grant 61833010 and in part by the Fundamental Research Funds for the Central Universities. (Corresponding author: Wei Cui)}

\author[SCUT]{Beili Gong}\ead{aublgong@mail.scut.edu.cn},    
\author[UNSW]{Daoyi Dong}\ead{daoyidong@gmail.com},               
\author[SCUT]{Weizhou Su}\ead{wzhsu@scut.edu.cn},  
\author[SCUT]{Wei Cui}\ead{aucuiwei@scut.edu.cn} 

\address[SCUT]{School of Automation Science and Engineering, South China University of Technology, Guangzhou 510641, China}  
\address[UNSW]{School of Engineering and Information Technology, University of New South Wales, Canberra, ACT 2600, Australia}             

\begin{abstract}
This paper investigates waveform estimation (tracking) of the time-varying force in a two-level optomechanical system with backaction noise by Kalman filtering.
It is assumed that the backaction and measurement noises are Gaussian and white.
By discretizing the continuous-time optomechanical system, the state of the resulting system can be estimated by the unbiased minimum variance Kalman filtering.
Then an estimator of the time-varying force is obtained, provided that the external force is also in discrete time.
Furthermore, the accuracy of the force estimation, described by the mean squared error, is derived theoretically.
Finally, the feasibility of the proposed algorithm is illustrated by comparing the theoretical accuracy with the numerical accuracy in a numerical example.
\end{abstract}


\end{frontmatter}

\section{Introduction}
Quantum cavity optomechanics, coupling the optical field to the mechanical resonator by radiation pressure or photothermal force, has been widely investigated in the past decade \cite{Dorsel:1983,Milburn:2011,BAspelmeyer:2014,Guolong:2018,Aspelmeyer:2012}.
It is not only a platform for investigating the fundamental questions on the quantum behavior of macroscopic systems \cite{Zheng:2016,Buchmann:2012}, but also a novel quantum device for high precision measurements \cite{Bariani:2013,Arvanitaki:2013,Heurs:2018}.
Moreover, an optomechanical system can be used as an optomechanical force sensor to measure external force, which is also called stochastic force as a sum of thermal noise and external signal.
Over the last few decades, many endeavors have been taken to study force estimation for optomechanical force sensors \cite{Tsang2011Fundamental,Taylor:2014,Armata:2017,Branford:2018,Zhang:2017Optomechanical}.
However, the majority of the results focus on the limit of estimation accuracy, instead of providing an estimator for external force.

In recent years, how to obtain an estimator of external force in quantum optomechanical systems has attracted many attentions.
In a quantum-enhanced interferometer, the optomechanical motion and force measurements have been demonstrated experimentally \cite{Iwasawa:2013}.
Based on the measurement, the estimation of the external stochastic force has been achieved by optical phase tracking and quantum smoothing techniques.
A statistical framework for the problem of parameter estimation from a noisy optomechanical system has been proposed \cite{Tsang2013Optomechanical}.
In this framework, three algorithms, namely, averaging algorithm, radiometer algorithm, and expectation maximization algorithm, have been applied to obtain an estimator of the noise power of the external stochastic force.
However, the algorithms become unavailable when the external force to be estimated is deterministic but time-varying.
In an optomechanical force sensor, the backaction noise introduced by quantum radiation-pressure fluctuations inevitably affects the dynamics of quantum system and further influences force estimation \cite{Tsang2010,Wimmer:2014,Bariani:2015,Yanay:2016}.
In aforementioned papers, the effects of backaction noise are not taken into account in force estimation.
That is, the estimation of time-varying force in quantum cavity optomechanics with backaction noise is still an open problem.

Note that the dynamic of a quantum optomechanical system can be described by a dissipationless linear Gaussian equation under quantum nondemolition measurements \cite{Tsang:2013metrology,Tsang2011Quantum,Mcmillen:2017}.
Inspired by the applications of Kalman filter in quantum sensors, e.g., estimating the phase of a light beam \cite{Yonezawa:2012}, estimating the waveform in a paradigmatic atomic sensor \cite{Ricardo:2018}, and estimating the quantum state of an optomechanical oscillator in real time \cite{Wieczorek:2015,Genoni:2017},
it is natural to apply the filtering theory on linear Gaussian systems to obtain an estimator of time-varying force.
This paper aims at proposing an algorithm to estimate the time-varying force for a quantum optomechanical system with backaction noise.
According to the Heisenberg equation, the evolution of quantum optomechanics with a two-level system is described by a linear stochastic equation.
Suppose that all noises of backaction and measurement are Gaussian and white \cite{Ricardo:2018,Trees:1968,Trees:1971}.
The measurement output is assumed discrete in time.
We convert the continuous-time system into a discrete-time linear Gaussian system.
Then the problem of the force estimation of an optomechanical system becomes a problem of input estimation of a linear Gaussian system.
Subsequently, the unbiased minimum variance Kalman filtering is applied to achieve an unbiased estimate of the system state, albeit the time-varying force is unknown.
Based on the estimated state, an unbiased estimator of the external force is obtained.
The theoretical accuracy, i.e., the mean squared error, of the force estimator is given as well.
Finally, an example is proposed to demonstrate the feasibility of the estimation method.

This paper is organized as follows. In Section \ref{sec:model}, a brief introduction to optomechanical systems and a stochastic differential equation are presented.
In Section \ref{estimation}, we consider the estimation of the time-varying force for a quantum optomechanical system.
Based on the unbiased minimum variance Kalman filtering, the estimators of the system state and the external force are presented and the theoretical accuracy of the force estimator is derived.
This section also considers an example illustrating the force estimation.
We summarize our conclusion in Section \ref{sec:conclusion}.

\section*{Notation}

A matrix $M\in \mathbb{R}^n$ is positive definite (semi-definite) if for any $\mu  \in \mathbb{R}^n$, such that $\mu^T  M {\mu} > 0$~$(\mu^T M{\mu} \ge 0)$, where ${\mu}^T$  denotes the conjugate transpose of $\mu$.
Write an $n \times n$ matrix $C=\mathrm{diag}(a_1, \dots, a_n)$ for a diagonal matrix whose diagonal entries are $a_1, \dots, a_n$.

\section{Model}\label{sec:model}
\begin{figure}[htb]
  \centering
  \includegraphics[width=7cm]{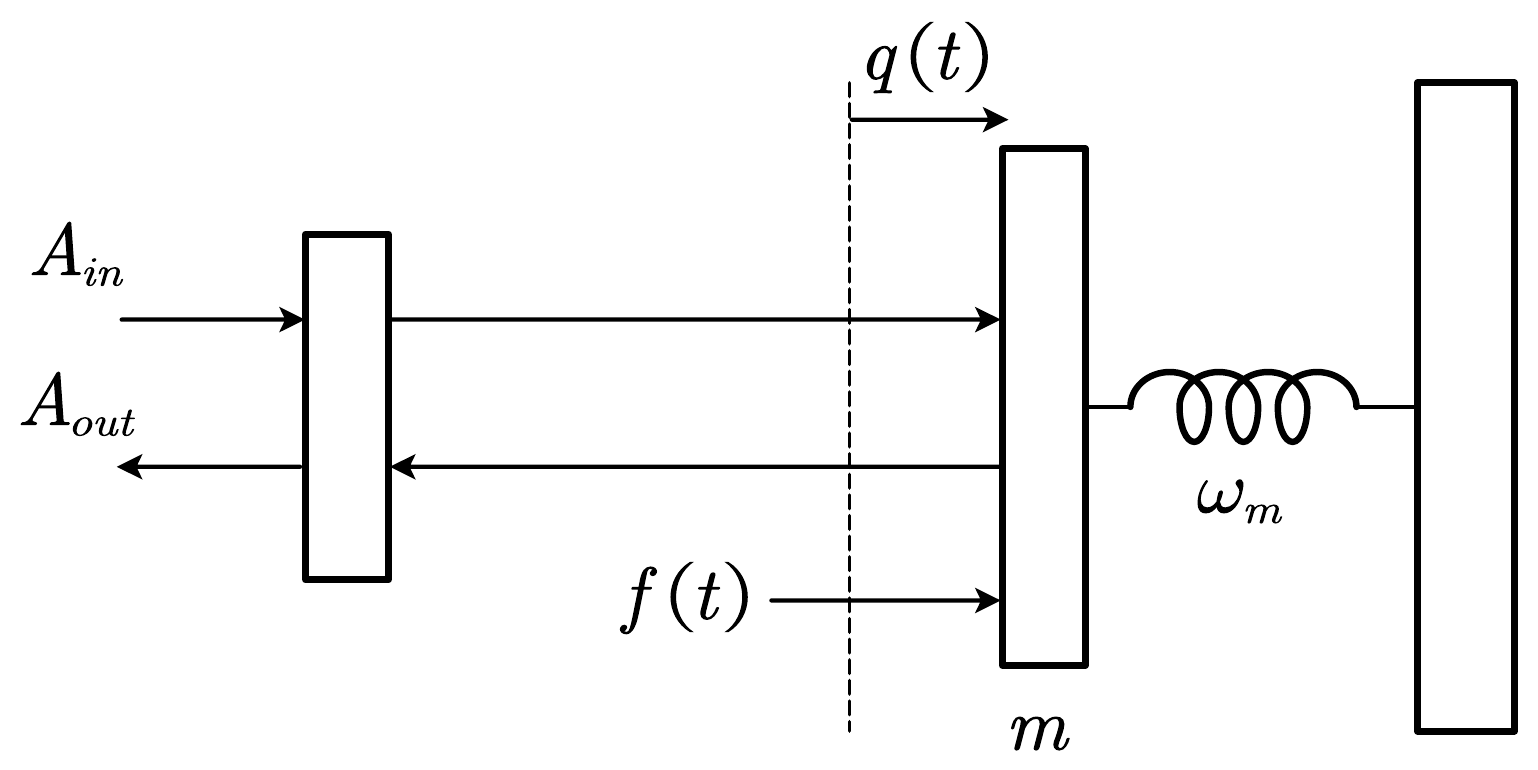}
   \caption{Schematic of optomechanical force sensor.}\label{Fig:quantum_optomechanical_sensor}
\end{figure}

Consider an optomechanical force sensor, as depicted in Fig.~ \ref{Fig:quantum_optomechanical_sensor}, in which a quantum harmonic oscillator is coupled to a Fabry-P\'{e}rot cavity with a moving mirror \cite{Tsang:2013metrology,Tsang:2012}.
On the left of the partial transmission mirror, the optical cavity is pumped on-resonance with the input beam $A_{in}$, and $A_{out}$ represents the output beam.
The mirror is simulated optically as perfectly reflecting and mechanically as a quantum harmonic oscillator with position operator $q(t)$, momentum operator $p(t)$, mass $m$, and resonant frequency $\omega_m$ \cite{Tsang2011Fundamental,Tsang2010,Tsang2011Quantum}.
The signal $f(t)$ is the external force acting on the mirror.
This framework is a basic model for more complex optomechanical force detectors with configurations, such as complex optical and mechanical mode structures and detuning cavity excitation.

Generally speaking, $A_{out}$ is measured to infer whether or not $f(t)$ disturbs the movement of the mirror in optomechanical force sensor.
Based on input-output formalism, the force detection sensitivity has been investigated; see, for example, \cite{Taylor:2014,Huang:2017}.
Since the motion of the mirror is observable, one can also extract information about the force by continuously measuring the position of mirror.
It is assumed that the optical cavity dynamics can be adiabatically eliminated and the optomechanical system experiences no dissipation.
Subsequently, the Hamiltonian of the optomechanical system is written as \cite{Tsang2011Fundamental,Tsang2011Quantum}
\begin{eqnarray*}
H(t) = \frac{{p^2(t)}}{2m} + \frac{m\omega _m^2 {q^2(t)}}{2} - q(t)f(t),
\end{eqnarray*}
where $f(t)$ is the time-varying force to be estimated, $p(t)$ and $q(t)$ satisfy the canonical commutation relation $\left[ {q(t),p(t)} \right] = i\hbar$ with $\hbar=1$.
Under quantum nondemolition position measurement, the backaction noise ${\zeta(t)}$ acts on the momentum only.
Accordingly, the Heisenberg equation of single motion can be described by \cite{Tsang2011Fundamental,Tsang2013Optomechanical}
\begin{eqnarray}\label{Heisenberg_equation}
\begin{aligned}
  \frac{d q(t)}{dt} &= \frac{ p(t)}{m},\\
\frac{d p(t)}{dt} &=  - m\omega _m^2 q(t) + f(t) + \zeta(t),
\end{aligned}
\end{eqnarray}
and the observation process can be given by
\begin{eqnarray}\label{observation}
y(t) = q(t) + \eta(t),
\end{eqnarray}
where $\eta(t)$ is measurement noise.

Let $x(t) = {[{\begin{matrix}q(t)&p(t)\end{matrix}}]^T}$ be the state of the optomechanical system and ${\xi}(t) = {[{\begin{matrix}0&{\zeta }(t)\end{matrix}}]^T}$ be the backaction noise vector.
The system evolution \eqref{Heisenberg_equation} with the measurement output \eqref{observation} can be rewritten as a stochastic differential equations, i.e,
\begin{eqnarray}\label{continuous_equation}
\begin{aligned}
\frac{dx(t)}{dt} &= {A_0}x(t) + {B_0}f(t) + {\xi}(t),\\
y(t) &= {H_0}x(t) + \eta(t).
\end{aligned}
\end{eqnarray}
Here, the matrices $A_0$, $B_0$ and $H_0$ are given by
\begin{eqnarray*}
A_0 =\left[\begin{matrix}
  0&{1/m} \\
 - m\omega _m^2&0
\end{matrix}\right], B_0 = \left[\begin{matrix}0\\1\end{matrix}\right], H_0= [\begin{matrix}1&0\end{matrix}].
\end{eqnarray*}
Note that $B_0$ has full column rank.

\begin{assum}
The measurement noise and backaction noise are assumed to be zero-mean Gaussian and white \cite{Ricardo:2018,Trees:1968,Trees:1971},
such that $E[\zeta(t){\zeta(t)^{T}}(t')]= D_0\delta(t-t')$ and
\begin{eqnarray*}\nonumber
 E[\xi(t){\xi^{T}}(t')] &=& Q_0\delta(t-t'),\\
 E[\eta(t){\eta^{T}}(t')]&=& R_0\delta(t-t'),
\end{eqnarray*}
where $D_0$ and $R_0$ are positive definite, $Q_0 = \mathrm{diag}(0,D_0)$ is positive semi-definite, and $\delta(t)$ is the Dirac delta function.
\end{assum}

Based on the above settings, the optomechanical system given by \eqref{continuous_equation} can be considered as a linear Gaussian system, and the external force $f(t)$ in \eqref{Heisenberg_equation} can be regarded as the input of the system.
In this case, the estimation of the external force can be converted into an estimation problem of the input for a linear Gaussian system.

\section{Force Estimation and Accuracy}\label{estimation}
A key task in parameter estimation is to maximize estimation accuracy and obtain an estimator that approaches to the limit of estimation accuracy \cite{Iwasawa:2013,Paris2009,Pasquale:2013}.
When an estimation problem related to Gaussian systems, the efficient methods are the well-known Kalman filtering and its extensions \cite{Ricardo:2018,Liu:2015,Setter:2018}, which are real-time algorithms to obtain the state estimate, provided that the input of the systems is known.
However, it is hard to estimate the state in continuous-time linear system when the input is time-varying and unknown, not to mention constructing an estimator for the unknown input only by measurement output.
Inspired by the autoregressive input estimation using robust two-stage Kalman filtering in \cite{zhang:2016},
we can first obtain the state estimate by the unbiased minimum variance Kalman filtering based on the measurement output, and then construct an estimator for the external force according to the relationship between the system state and the external force.
In this section, an unbiased estimator for the external force is explicitly given and the theoretical accuracy of the estimated result is derived.

\subsection{Discretization}
Note that the Kalman-Bucy filter is proposed to solve the state estimation problem in continuous-time \cite{Ricardo:2018,Kailath1968}, and this filter works well only if the input dynamics and the statistical information of the noise are known.
In other words, the Kalman-Bucy filter becomes invalid when the dynamics of the input signal are not provided.
Since the measurement signals are discrete in time when obtaining by some digital devices, it is reasonable to convert an estimation problem from continuous time to discrete time.
With assuming zero-order hold for the input and continuous integration for the measurement and backaction noises,
the continuous-time state-space model \eqref{continuous_equation} can be discretized to
\begin{eqnarray}\label{linear_equation}
\begin{aligned}
x_{k+1}&= Ax_{k} + Bf_{k} + w_{k},\\
y_{k} &= Hx_{k} + v_{k},
\end{aligned}
\end{eqnarray}
with the sampling period $\Delta t$.
In this discrete-time state-space equation, $x_{k}$ represents the system state at time $k$, $f_{k}$ represents the system input which may be deterministic or stochastic, and $y_{k}$ is the measurement output.
Also, $w_{k}, v_{k}$ are sequences of white noise with zero mean and covariance $Q_k$ and $R_k$, respectively.
Here the system matrices $(A,B,H,Q_k,R_k)$ are given in terms of $(A_0,B_0,H_0,Q_0,R_0)$ and ${\Delta t}$:
\begin{eqnarray*}
A &= {e^{{A_0}\Delta t}};B = \int_{\tau  = 0}^{\Delta t} {{e^{{A_0}\tau }}d\tau } {B_0};H = {H_0},\\
{Q_k} &= \int_{\tau  = 0}^{\Delta t} {{e^{{A_0}\tau }}{Q_0}{e^{A_0^T\tau }}d\tau };{R_k} = {R_0}/{\Delta t},
\end{eqnarray*}
where $A_0^T$ is the transpose of $A_0$.

\subsection{Force Estimation}
One of the well-known algorithms to solve the unknown input dynamics problem is the unbiased minimum variance Kalman filtering, in which the input of the system is treated as a disturbance \cite{zhang:2016,Kitanidis:1987,Keller:1997,Kim:2006}.
\begin{lem}\label{lem:UMVF}
(Unbiased Minimum Variance Kalman Filtering) Consider a linear stochastic system described by \eqref{linear_equation} with the input $f_k$ being unknown, the unbiased minimum-variance state estimate is given by \cite{Kitanidis:1987}
\begin{eqnarray}\label{state_filter}
{\hat x_{k + 1|k + 1}} = A{\hat x_{k|k}}+ L_{k + 1}\left[{y_{k+1}}- HA{\hat x_{k|k}} \right],
\end{eqnarray}
where $L_{k + 1}$ is the Kalman gain given by
\begin{eqnarray}\nonumber
\begin{aligned}
L_{k + 1}&=P_{k + 1|k}H^T{C_{k+1}^{ - 1}}\\
&\hspace{0.3cm}+[{B - P_{k + 1|k}H^T{C_{k+1}^{ - 1}}HB}]\\
&\hspace{0.3cm}\times[ {{B^T}{H^T}{{C_{k+1}^{ - 1}}}HB}]^{ - 1}B^TH^T{C_{k+1}^{ - 1}}
\end{aligned}
\end{eqnarray}
with
\begin{eqnarray} \nonumber
\begin{aligned}
P_{k + 1|k}&= AP_{k|k}A^T + Q_k,\\
C_{k + 1} &= HP_{k + 1|k}H^T + R_k,\\
P_{k + 1|k + 1} &= P_{k + 1|k} - P_{k + 1|k}H^T\times{C_{k+1}^{ - 1}}HP_{k + 1|k}\\
 &\hspace{0.3cm}+[B - P_{k + 1|k}H^T{C_{k+1}^{ - 1}}HB]\\
 &\hspace{0.3cm}\times[ B^TH^T{C_{k+1}^{ - 1}}HB]^{ - 1}\\
 &\hspace{0.3cm}\times [B - P_{k + 1|k}H^T{C_{k+1}^{ - 1}}HB]^T.
\end{aligned}
\end{eqnarray}
\end{lem}
Here $P_{k|k}$ is referred to as the error covariance matrix that
\begin{eqnarray}\label{ex_P}
 P_{k|k}=E[({x_{k}-\hat x_{k|k}})(x_{k}-{\hat x_{k|k}})^T].
\end{eqnarray}

Although there is an unknown input with untraceable property, such filtering is still capable of achieving unbiased state estimates.
Moreover, this filtering provides an accurate state estimate to maintain the trait of being unbiased $E[{\hat x_{k|k}}] = E[x_{k}]$, which has been proved in \cite{Kitanidis:1987}.

\begin{thm}\label{thm:hat_f}
Since $B$ have full column rank, an unbiased estimator of the external force can be given by
\begin{eqnarray}\label{force_esitimator}
 \hat f_{k} = B^{+}[{\hat x_{k + 1|k + 1}} - A{\hat x_{k|k}}],
\end{eqnarray}
where $B^{+}$ represents the Moore-Penrose inverse of the matrix $B$, i.e., ${B^{+} }=(B^{T}B)^{-1}B^{T}$.
\end{thm}

\begin{pf}
According to Lemma \ref{lem:UMVF}, the state estimate can be obtained.
Together with the discrete-time state-space equation \eqref{linear_equation}, one can obtain an estimator of the external force from
\begin{eqnarray}\nonumber
Bf_{k} + \bar w_{k} = \hat x_{k + 1| k + 1} - A\hat x_{k|k},
\end{eqnarray}
where $\bar w_k$ consists of the driven noise and the innovation.
Note that $\bar w_{k}$ is white and $B$ is full column rank. It is clear that \eqref{force_esitimator} is an unbiased estimator of the force $f_k$.
A similar derivation can be found in \cite{zhang:2016}.
\end{pf}

\subsection{Estimation Accuracy}
The accuracy of an estimator is used to quantify the difference between the estimated values and what is estimated.
A well-known performance index of the accuracy is the mean squared error, which measures the average of the squares of the errors.
The mean squared error of an estimator $\hat \theta$ with respect to an unknown parameter $\theta$ is defined as
\begin{eqnarray}\nonumber
  \varepsilon _{\theta}^2  \buildrel \Delta \over = E[({\hat \theta}-\theta)({\hat \theta}-\theta)^T].
\end{eqnarray}
From the definition, we have $\varepsilon _{x_{k}}^2=P_{k|k}$.
\begin{thm}\label{thm:V(f)}
The mean squared errors of the force and state estimators, namely $\varepsilon _{f_{k}}^2$ and $\varepsilon _{x_{k}}^2$, satisfy
\begin{eqnarray}\label{Eq:mse_force}
\begin{aligned}
\varepsilon _{f_{k}}^2 &=M_{k+1}HA\varepsilon_{x_{k}}^{2}A^{T}H^{T}{M_{k+1}^{T}}\\
&\hspace{0.3cm}+M_{k+1}HQ_kH^{T}{M_{k+1}^{T}}+M_{k+1}R_{k+1}{M_{k+1}^{T}},
\end{aligned}
\end{eqnarray}
where $M_{k+1}={B^+}L_{k + 1}$, and $Q_k$ and $R_{k+1}$ are, respectively, the variances of the backaction noise and  measurement noise in the discrete-time state-space equation \eqref{linear_equation}.
\end{thm}

\begin{pf}
According to the definition of mean squared error, it is clear that
\begin{eqnarray}\label{Eq:f^2}
 \varepsilon _{f_{k}}^2 = E[\hat{f}_k^2]-E[{\hat{f}_k}f_k^T]-E[f_k\hat{f}_k^T] + E[{f_k}^2].
\end{eqnarray}
Substituting \eqref{state_filter} into \eqref{force_esitimator} yields
\begin{eqnarray*}
\begin{aligned}
 \hat f_{k} = B^{+}L_{k + 1}({y_{k+1}} - HA{\hat x_{k|k}}).
\end{aligned}
\end{eqnarray*}
Define $M_{k+1}={B^+}L_{k + 1}$ for convenience.
From \eqref{linear_equation}, $\hat{f}_k$ can be further expressed as
\begin{eqnarray}\label{hat_f}
\begin{aligned}
\hat{f}_k = M_{k+1}[HA(x_{k}-\hat x_{k|k})+HBf_k+Hw_k+v_{k+1}].
\end{aligned}
\end{eqnarray}
Taking expectation on the both sides of \eqref{hat_f} shows
\begin{eqnarray*}
\begin{aligned}
E[\hat{f}_k]& = E[M_{k+1}(HA(x_{k}-\hat x_{k|k})+HBf_k \\
&\hspace{4.8cm}+Hw_k+v_{k+1})]\\
& = M_{k+1}HE[A(x_{k}-\hat x_{k|k})]+M_{k+1}HBE[f_k],
\end{aligned}
\end{eqnarray*}
since $E[w_k]=0$ and $E[v_{k+1}]=0$.

Due to \eqref{hat_f}, $\hat{f}_{k}^{2}$ can be written as
\begin{eqnarray*}
\begin{aligned}
\hat{f}_{k}^{2}&=\hat{f}_k\hat{f}_k^{T}\\
&=M_{k+1}[HA(x_{k}-\hat x_{k|k})+HBf_k+Hw_k+v_{k+1}]\\
&\hspace{0.1cm}\times [HA(x_{k}-\hat x_{k|k})+HBf_k+Hw_k+v_{k+1}]^{T}M_{k+1}^{T}\\
&=M_{k+1}(N_1+N_2+N_3+N_4)M_{k+1}^{T},
\end{aligned}
\end{eqnarray*}
where
\begin{eqnarray*}
\begin{aligned}
N_1&=HA(x_{k}-\hat x_{k|k})(x_{k}-\hat x_{k|k})^{T}A^{T}H^{T}\\
&\hspace{0.3cm}+HA(x_{k}-\hat x_{k|k})f_k^{T}B^{T}H^{T}\\
&\hspace{0.3cm}+HA(x_{k}-\hat x_{k|k})w_k^{T}H^{T}+HA(x_{k}-\hat x_{k|k})v_{k+1}^{T},\\
N_2&=HBf_k(x_{k}-\hat x_{k|k})^{T}A^{T}H^{T}+HBf_kf_k^{T}B^{T}H^{T}\\
&\hspace{0.3cm}+HBf_kw_k^{T}H^{T}+HBf_kv_{k+1}^{T},\\
N_3&=Hw_k(x_{k}-\hat x_{k|k})^{T}A^{T}H^{T}+Hw_kf_k^{T}B^{T}H^{T}\\
&\hspace{0.3cm}+Hw_kw_k^{T}H^{T}+Hw_kv_{k+1}^{T},\\
N_4&=v_{k+1}(x_{k}-\hat x_{k|k})^{T}A^{T}H^{T}+v_{k+1}f_k^{T}B^{T}H^{T}\\
&\hspace{0.3cm}+v_{k+1}w_k^{T}H^{T}+v_{k+1}v_{k+1}^{T}.
\end{aligned}
\end{eqnarray*}
Note that the estimators of the system state and the external force are unbiased.
One can obtain the expectations of $N_1, N_2, N_3, N_4$, respectively,
\begin{eqnarray*}
\begin{aligned}
E[N_1]&=HAE[(x_{k}-\hat x_{k|k})(x_{k}-\hat x_{k|k})^{T}]A^{T}H^{T},\\
E[N_2]&=HBE[f_kf_k^{T}]B^{T}H^{T},\\
E[N_3]&=HE[w_kw_k^{T}]H^{T},\\
E[N_4]&=E[v_{k+1}v_{k+1}^{T}],
\end{aligned}
\end{eqnarray*}
since $f_k,w_k,v_{k+1}$ are uncorrelated with $x_{k}$ and $\hat x_{k|k}$ at time $k$.
Therefore, the expectation of $\hat{f}_{k}^{2}$ is
\begin{eqnarray}\label{Eq:E[f^2]}
\begin{aligned}
E[\hat{f}_{k}^{2}]&=M_{k+1}HA\varepsilon _{x_{k}}^2A^{T}H^{T}M_{k+1}^{T}+E[f_k^{2}]\\
&\hspace{0.3cm}+M_{k+1}HQ_kM_{k+1}^{T}+M_{k+1}R_{k+1}M_{k+1}^{T},
\end{aligned}
\end{eqnarray}
which follows from the definition of mean squared error of $\hat x_{k|k}$ and the fact $M_{k+1}HB=I$.

Additionally, it can be shown that
\begin{eqnarray}\label{Eq:f_hatf}
  E[{\hat{f}_k}f_k^T] = E[f_k\hat{f}_k^T] = E[{f_k}^2],
\end{eqnarray}
since $x_{k},\hat x_{k|k}, w_k, v_{k+1}$ are uncorrelated with $f_k$ at time $k$ and $M_{k+1}HB=I$.

After substituting \eqref{Eq:E[f^2]} and \eqref{Eq:f_hatf} into \eqref{Eq:f^2}, the mean squared error of the force estimator becomes
\begin{eqnarray*}
\begin{aligned}
\varepsilon _{f_{k}}^2 &=M_{k+1}HA\varepsilon_{x_{k}}^{2}A^{T}H^{T}{M_{k+1}^{T}}\\
&\hspace{0.3cm}+M_{k+1}HQ_kH^{T}{M_{k+1}^{T}}+M_{k+1}R_{k+1}{M_{k+1}^{T}}.
\end{aligned}
\end{eqnarray*}
The proof is then completed.
\end{pf}

Theorem \ref{thm:V(f)} analytically provides the estimation accuracy of the force in the discrete time.
Note that $\varepsilon _{f_{k}}^2$ is a function of $\varepsilon _{x_{k}}^2$ and is lower-bounded by $M_{k+1}HQ_kH^{T}{M_{k+1}^{T}}+M_{k+1}R_{k+1}{M_{k+1}^{T}}$.

\subsection{Numerical Example}
In this part, a numerical example is used to demonstrate the proposed method.
Assume that the external force to be estimated is time-varying and obeys the Gaussian distribution
with unit-mean and the variance being $0.5$.
Consider a quantum optomechanical system with its parameters $m = 5.88\times10^{-4}\mathrm{kg},~\omega_m = 1.76\times10^5\mathrm{rad/s}$.
Let the initial state of system be $x(0)=[q(0)~~p(0)]^T$ with $q(0)=10^{-6}\mathrm{m}$ and $p(0)=10^{-6}\mathrm{kgm/s}$, and the sampling period $\Delta t$ be $10^{-4}\mathrm{s}$.
The variance matrices of all noises are $Q_0 = \mathrm{diag}(0,D)$ and $R_k = D$, where $D=10^{-14}$.

\begin{figure}[htb]
  \centering
  \includegraphics[width=8cm]{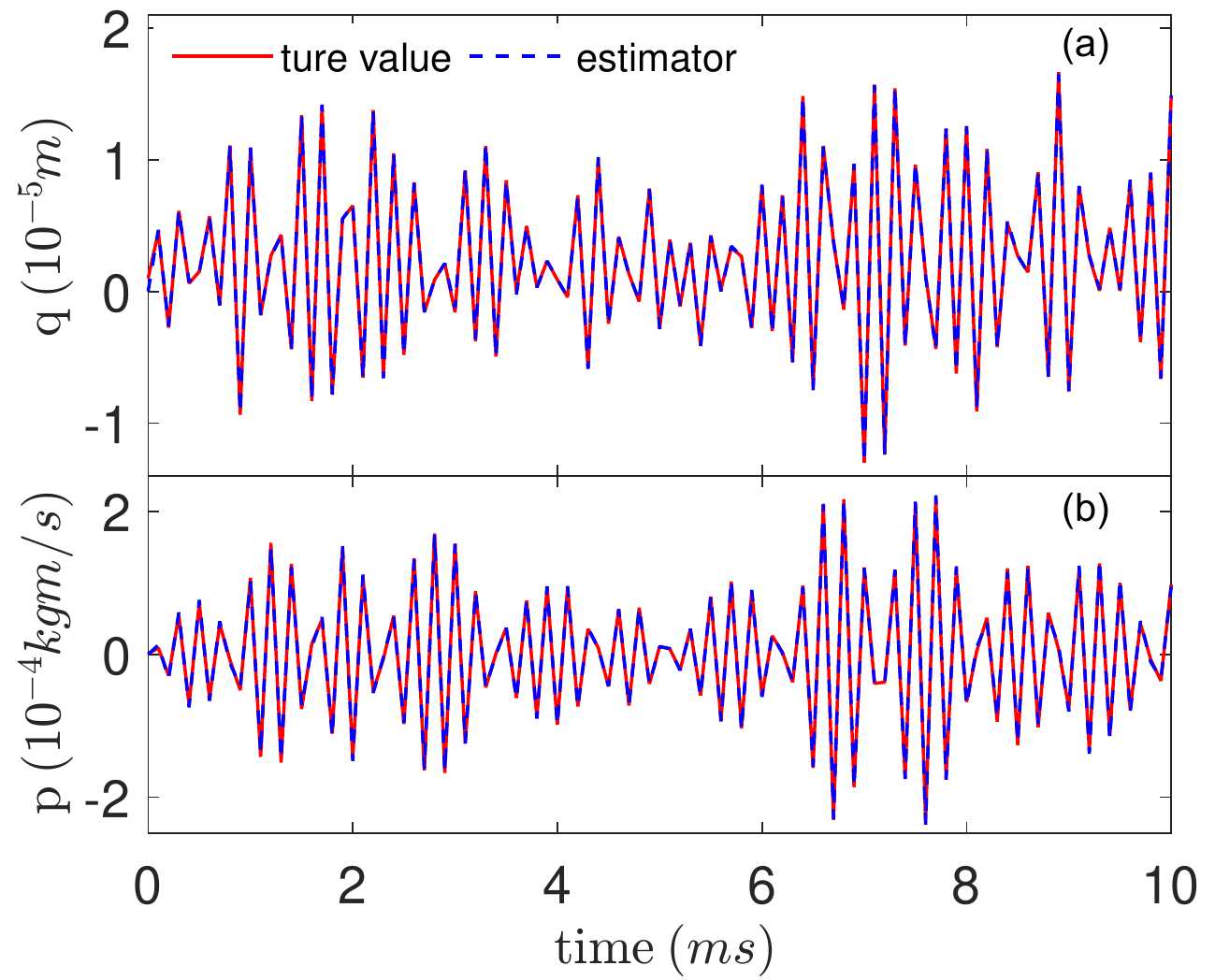}
  \caption{The true and the estimated values of the system state: (a) position and (b) momentum.}\label{fig:pq_filter}
\end{figure}

Based on the measurement outputs, the state estimate is obtained by Lemma \ref{lem:UMVF}. The estimated position and momentum are shown in Fig.~\ref{fig:pq_filter}, wherein the estimated values are close to the true values with small errors, indicating that the method of estimating the system state is feasible.

\begin{figure}[htb]
  \centering
  \includegraphics[width=8cm]{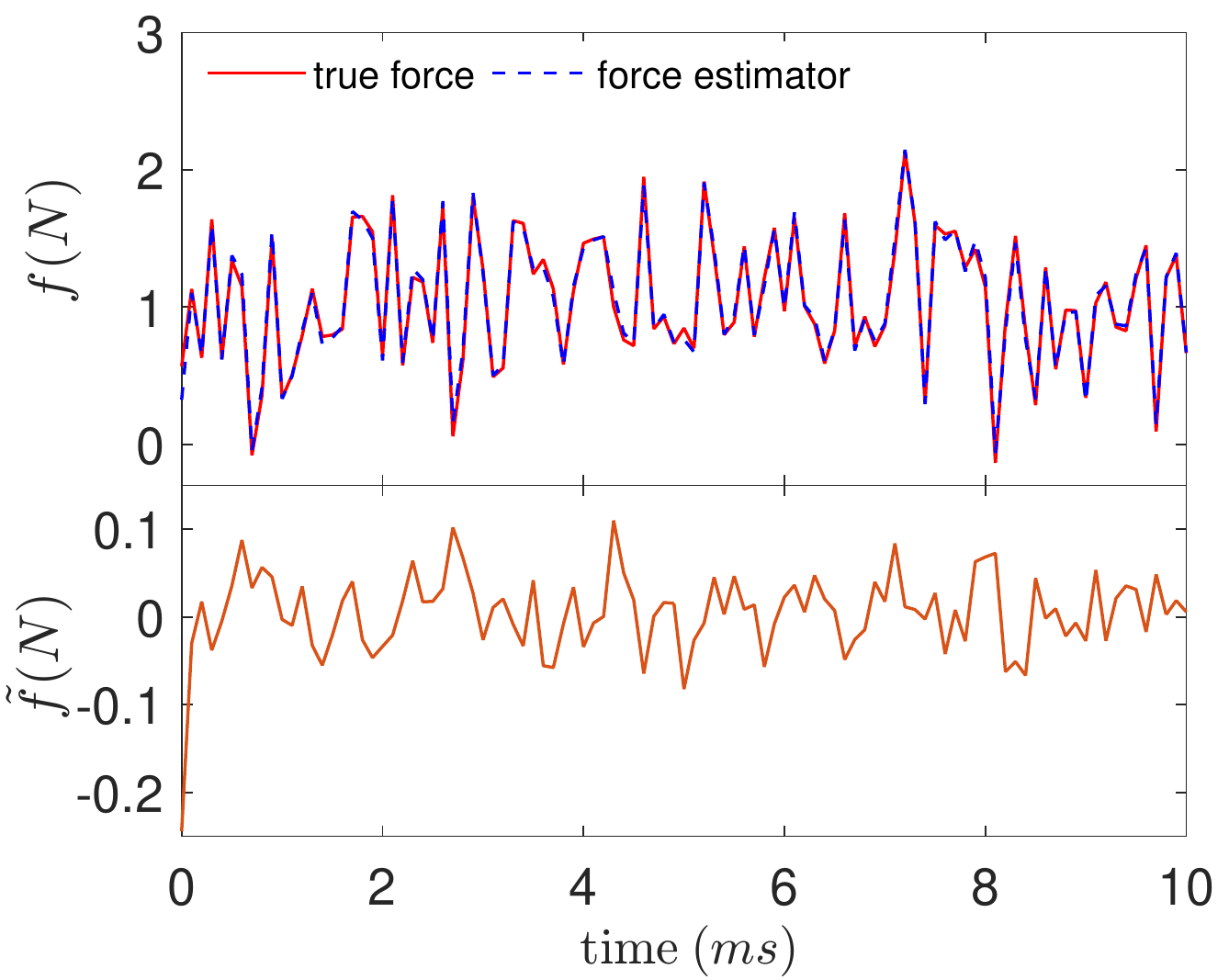}
  \caption{The trajectories of the true force and force estimator, and the error between them.}\label{fig:force_estimator}
\end{figure}

Accordingly, an estimator of the time-varying force is obtained by Theorem \ref{thm:hat_f}.
On the top half of Fig.~\ref{fig:force_estimator}, the red solid curve represents the true force and the blue dash curve is the force estimator in a single experiment.
The orange curve in the bottom of Fig.~\ref{fig:force_estimator} shows the error $\tilde f$ ($\tilde{f_k} = {\hat f}_k-{f_k}$ for each time $k$) between the true force and the estimated force.
Note that the system in the numerical example is obviously of ergodicity.
Define $E_T$ as an operator which takes average over time.
Through simple computation, the time average of $\tilde{f}$ is given by $E_T(\tilde{f}) = 4.7574\times10^{-7}$,
which reveals the unbiasedness of the force estimator and illustrates the effectiveness of estimation algorithm of the time-varying force proposed in Theorem \ref{thm:hat_f}.

\begin{figure}[htb]
  \centering
  \includegraphics[width=8cm]{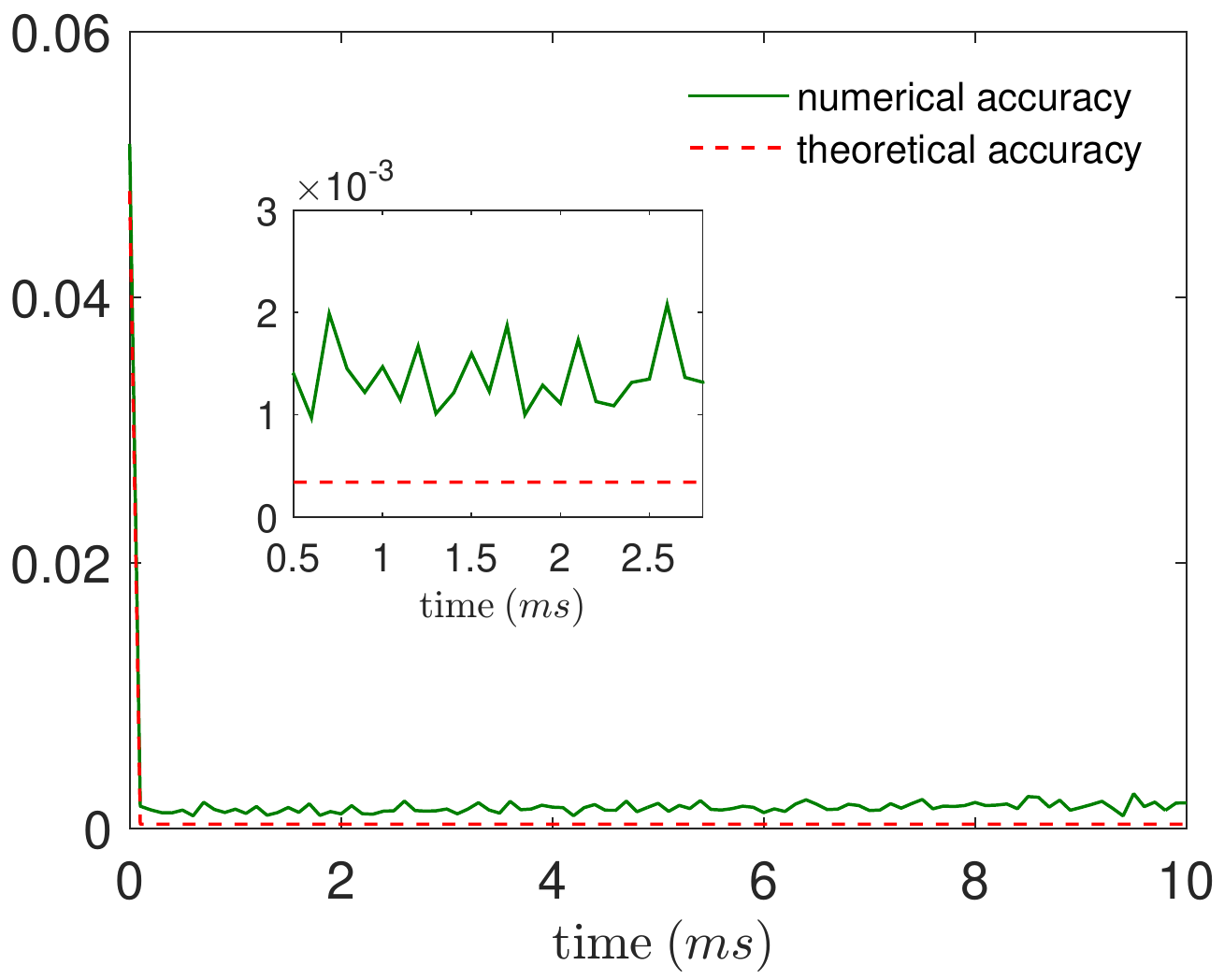}
  \caption{The theoretical and numerical accuracy of force estimation with $N_M=100$.}\label{fig:accuracy}
\end{figure}

It is worth calculating the accuracy of the estimation through a large number of experiments.
Define the numerical accuracy \cite{Liu:2015} as
\begin{eqnarray}\label{Eq:numerical_accuracy}
 {V_{N}}(f_k) = \frac{1}{{{N_M}}}\sum\limits_{i = 1}^{{N_M}} {{{\left( {f_k^i - \hat f_k^i} \right)}^2}}
\end{eqnarray}
where $\hat f_k^i$ and $f_k^i$ are, respectively, the estimated value and the true value in the $i$th simulation experiment, and $N_M$ is the total number of experiments.
The numerical accuracy and the theoretical accuracy with respect to the time index are depicted in Fig.~\ref{fig:accuracy} with $N_M=100$.
It shows that the numerical accuracy is always larger than the theoretical accuracy, as a simple implication of Theorem \ref{thm:V(f)}.
Combining Fig.~\ref{fig:force_estimator} with Fig.~\ref{fig:accuracy} indicates that the proposed algorithm for estimating the time-varying force in an optomechanical system is effective and feasible.

\section{Conclusion}\label{sec:conclusion}
In this paper, we have investigated the estimation problem of an time-varying force in a two-level quantum cavity optomechanics with backaction noise.
Based on the Heisenberg equation and the observation process, the evolution of an ideal system, eliminating adiabatically any optical cavity dynamics and neglecting intrinsic mechanical losses, has been described by a linear stochastic differential equation.
The estimation of the external force for the optomechanical system can be converted into the input estimation for a linear Gaussian system.
Using the unbiased minimum variance Kalman filtering, we have given the state estimate whenever external force is unknown.
Then an estimator for the external force has been constructed and the theoretical accuracy of the force estimation has been derived according to the relationship between the system state and the external force.
Finally, a numerical example has illustrated the force estimation and demonstrated the feasibility of the estimation algorithm, by comparing the numerical accuracy with the theoretical accuracy.


\begin{thebibliography}{1}
\bibitem{Dorsel:1983}
A.~Dorsel, J.~D.~McCullen, P.~Meystre, E.~Vignes, and H.~Walther, ``Optical bistability and mirror confinement induced by radiation pressure," \emph{Phys. Rev. Lett.}, vol.~51, no.~17, pp.~1550--1553, Oct. 1983.

\bibitem{Milburn:2011}
G.~J.~Milburn and M.~J.~Woolley, An introduction to quantum optomechanics, \emph{Acta. Physica. Slovaca.},
vol.~61, no.~5, pp.~483--601, Oct. 2011.

\bibitem{BAspelmeyer:2014}
M.~Aspelmeyer, T.~J.~Kippenberg, and F.~Marquardt, \emph{Cavity Optomechanics.} Berlin: Springer-Verlag, 2014.

\bibitem{Guolong:2018}
G.~Li and X.~Wang, ``Heisenberg-limited estimation of the coupling rate in an optomechanical system with a two-level system," \emph{Phys. Rev. A}, vol.~98, no.~1, pp.~013803, Jul. 2018.

\bibitem{Aspelmeyer:2012}
M.~Aspelmeyer, P.~Meystre, and K.~Schwab, ``Quantum optomechanics," \emph{Phys. Today}, vol.~65, no.~7, pp.~29, Jul. 2012.

\bibitem{Zheng:2016}
Q.~Zheng, Y.~Yao, and Y.~Li, ``Optimal quantum parameter estimation in a pulsed quantum optomechanical system," \emph{Phys. Rev. A}, vol.~93, no.~1, pp.~013848, Jan. 2016.

\bibitem{Buchmann:2012}
L.~F.~Buchmann, L.~Zhang, A.~Chiruvelli, and P.~Meystre, ``Macroscopic tunneling of a membrane in an optomechanical double-well potential," \emph{Phys. Rev. Lett.}, vol.~108, no.~21, pp.~210403, May. 2012.

\bibitem{Bariani:2013}
H.~Tan, F.~Bariani, G.~Li, and P.~Meystre, ``Generation of macroscopic quantum superpositions of optomechanical oscillators by dissipation," \emph{Phys. Rev. A}, vol.~88, no.~2, pp.~023817, Aug. 2013.

\bibitem{Arvanitaki:2013}
A.~Arvanitaki and A.~A.~Geraci, ``Detecting high-frequency gravitational waves with optically levitated sensors," \emph{Phys. Rev. Lett.}, vol.~110, no.~7, pp.~071105, Feb. 2013.

\bibitem{Heurs:2018}
M.~Heurs, ``Gravitational wave detection using laser interferometry beyond the standard quantum limit," \emph{Philos. Trans. Royal Soc. A}, vol.~376, no.~2120, pp.~20170289, Apr. 2018.

\bibitem{Tsang2011Fundamental}
M.~Tsang, H.~M.~Wiseman, and C.~M.~Caves, ``Fundamental quantum limit to waveform estimation," \emph{Phys. Rev. Lett.}, vol.~106, no.~9, pp.~090401, Mar. 2011.


\bibitem{Taylor:2014}
X.~Xu and J.~M.~Taylor, ``Squeezing in a coupled two-mode optomechanical system for force sensing below the standard quantum limit," \emph{Phys. Rev. A}, vol.~90, no.~4, pp.~043848, Oct. 2014.

\bibitem{Armata:2017}
F.~Armata, L.~Latmiral, A.~D.~K.~Plato, and M.~S.~Kim, ``Quantum limits to gravity estimation with optomechanics," \emph{Phys. Rev. A}, vol.~96, no.~4, pp.~043824, Oct. 2017.

\bibitem{Branford:2018}
D.~Branford, H.~Miao, and A.~Datta, ``Fundamental quantum limits of multicarrier optomechanical sensors," \emph{Phys. Rev. Lett.}, vol.~121, no.~11, pp.~110505, Sep. 2018.

\bibitem{Zhang:2017Optomechanical}
W.~Zhang, Y.~Han,  B.~Xiong, and L.~Zhou, ``Optomechanical force sensor in non-Markovian regime," \emph{ New J.
Phys.}, vol.~19, no.~8, pp.~083022, Aug. 2017.

\bibitem{Iwasawa:2013}
K.~Iwasawa, K.~Makino, H.~Yonezawa, M.~Tsang, A.~Davidovic, E.~Huntington, and A.~Furusawa, ``Quantum-limited mirror-motion estimation," \emph{Phys. Rev. Lett.}, vol.~111, no.~16, pp.~163602, Oct. 2013.

\bibitem{Tsang2013Optomechanical}
S.~Z.~Ang, G.~I.~Harris, W.~P.~Bowen, and M.~Tsang, ``Optomechanical parameter estimation," \emph{New J. Phys.}, vol.~15, no.~10, pp.~103028, Oct. 2013.

\bibitem{Tsang2010}
M.~Tsang and C.~M.~Caves, ``Coherent quantum-noise cancellation for optomechanical sensors," \emph{Phys. Rev. Lett.},
vol.~105, no.~12, pp.~123601, Sep. 2010.

\bibitem{Wimmer:2014}
M.~H.~Wimmer, D.~Steinmeyer, K.~Hammerer, and M.~Heurs, ``Coherent cancellation of backaction noise in optomechanical force measurements," \emph{Phys. Rev. A}, vol.~89, no.~5, pp.~053836, May. 2014.

\bibitem{Bariani:2015}
F.~Bariani, H.~Seok, S.~Singh, M.~Vengalattore, and P.~Meystre, ``Atom-based coherent quantum-noise cancellation in optomechanics," \emph{Phys. Rev. A}, vol.~92, no.~4, pp.~043817, Oct. 2015.

\bibitem{Yanay:2016}
Y.~Yanay, J.~C.~Sankey, and A.~A.~Clerk, ``Quantum backaction and noise interference in asymmetric two-cavity optomechanical systems," \emph{Phys. Rev. A}, vol.~93, no.~6, pp.~063809, Jun. 2016.

\bibitem{Tsang:2013metrology}
M.~Tsang, ``Quantum metrology with open dynamical systems," \emph{New J. Phys.}, vol.~15, no.~7, pp.~073005, Jul. 2013.

\bibitem{Tsang2011Quantum}
M.~Tsang, ``Quantum backaction noise cancellation for linear systems," \emph{AIP Conference Proceedings}, Brisbane, Australia, Oct. 2011, vol.~1363, pp.~93--96.


\bibitem{Mcmillen:2017}
S.~McMillen, M.~Brunelli, M.~Carlesso, A.~Bassi, H.~Ulbricht, M.~G.~A.~Paris, and M.~Paternostro, ``Quantum-limited estimation of continuous spontaneous localization," \emph{Phys. Rev. A}, vol.~95, no.~1, pp.~012132, Jan. 2017.

\bibitem{Yonezawa:2012}
H.~Yonezawa, D.~Nakane, T.~A.~Wheatley, K.~Iwasawa, S.~Takeda, H.~Arao, K.~Ohki, K.~Tsumura, D.~W.~Berry, T.~C.~Ralph, H.~M.~Wiseman, E.~H.~Huntington, and A.~Furusawa, ``Quantum-enhanced optical-phase tracking,"
\emph{Science}, vol.~337, no.~6101, pp.~1514--1517, Sep. 2012.

\bibitem{Ricardo:2018}
R.~Jim\'{e}nez-Mart\'{\i}nez, J.~Ko{\l}ody\'{n}ski, C.~Troullinou, V.~G.~Lucivero, J.~Kong, and M.~W.~Mitchell,
``Signal tracking beyond the time resolution of an atomic sensor by Kalman filtering," \emph{Phys. Rev. Lett.}, vol.~120, no.~4, pp.~040503, Jan. 2018.

\bibitem{Wieczorek:2015}
W.~Wieczorek, S.~G.~Hofer, J.~Hoelscher-Obermaier, R.~Riedinger, K.~Hammerer, and M.~Aspelmeyer, ``Optimal state estimation for cavity optomechanical systems," \emph{Phys. Rev. Lett.}, vol.~114, no.~22, pp.~223601, Jun. 2015.

\bibitem{Genoni:2017}
M.~G.~Genoni, ``Cram\'{e}r-Rao bound for time-continuous measurements in linear Gaussian quantum systems,"  \emph{Phys. Rev. A}, vol.~95, no.~5, pp.~059908, May. 2017.

\bibitem{Trees:1968}
H.~L.~Van~Trees, \emph{Detection, Estimation and Modulation Theory: I}. \hskip 1em plus 0.5em minus 0.4em\relax New York: Wiley, 1968.

\bibitem{Trees:1971}
H.~L.~Van~Trees, \emph{Detection, Estimation and Modulation Theory: III. Radar-–Sonar Signal Processing and
Gaussian Signals in Noise}. \hskip 1em plus 0.5em minus 0.4em\relax New York: Wiley, 1971.

\bibitem{Tsang:2012}
M.~Tsang and R.~Nair, ``Fundamental quantum limits to waveform detection," \emph{Phys. Rev. A}, vol.~86, no.~4, pp.~042115, Oct. 2012.

\bibitem{Huang:2017}
S.~Huang and G.~S.~Agarwal, ``Robust force sensing for a free particle in a dissipative optomechanical system
with a parametric amplifier," \emph{Phys. Rev. A}, vol.~95, no.~2, pp.~023844, Feb. 2017.


\bibitem{Paris2009}
M.~G.~A.~Paris, ``Quantum estimation for quantum technology," \emph{Int. J. Quantum Inf.}, vol.~07, no.~supp01, pp.~125-137, 2009.

\bibitem{Pasquale:2013}
A.~D.~Pasquale, D.~Rossini, P.~Facchi, and V.~Giovannetti, ``Quantum parameter estimation affected by unitary disturbance,"  \emph{Phys. Rev. A}, vol.~88, no.~5, pp.~052117, Nov. 2013.

\bibitem{Liu:2015}
L.~Liu, B.~Qi, S.~Cheng, and Z.~Xi, ``High precision estimation of inertial rotation via the extended Kalman filter," \emph{Eur. Phys. J. D}, vol.~69, no.~11, pp.~261, Nov. 2015.

\bibitem{Setter:2018}
A.~Setter, M.~Toro\v{s}, J.~F.~Ralph, and H.~Ulbricht, ``Real-time Kalman filter: cooling of an optically levitated nanoparticle," \emph{Phys. Rev. A}, vol.~97, no.~3, pp.~033822, Mar. 2018.

\bibitem{zhang:2016}
H.~Zhuang and J.~Li, ``Robust two-stage Kalman filtering in presence of autoregressive input," in \emph{Proceedings of 2016 14th International Conference on Control, Automation, Robotics and Vision (ICARCV)}, Phuket, Thailand, Nov. 2016, pp.~1--6.

\bibitem{Kailath1968}
T.~Kailath, ``An innovations approach to least-squares estimation--Part I: Linear filtering in additive white noise,"  \emph{IEEE Trans. Automat. Control.}, vol.~13, no.~6, pp.~646--655, Dec. 1968.

\bibitem{Kitanidis:1987}
P.~K.~Kitanidis, ``Unbiased minimum-variance linear state estimation," \emph{Automatica}, vol.~23, no.~6, pp.~775--778, Nov.  1987.

\bibitem{Keller:1997}
J.~Y.~Keller and M.~Darouach, ``Optimal two-stage Kalman filter in the presence of random bias," \emph{Automatica},
vol.~33, no.~9, pp.~1745--1748, Sep. 1997.

\bibitem{Kim:2006}
K.~H.~Kim, J.~G.~Lee, and C.~G.~Park, ``Adaptive two-stage Kalman filter in the presence of unknown random bias,"
\emph{Int. J. Adapt. Control Signal Process}, vol.~20, no.~7, pp.~305--319, May. 2006.

\end{thebibliography}

\end{document}